\documentclass[12pt]{iopart} 
\input epsf
\usepackage{graphicx}
\begin{document}

\title[Electromigrated Au nanocontacts]{Structure and local charging of electromigrated Au nanocontacts}

\author{D~Arnold$^{1}$, M~Marz$^{1}$, S~Schneider$^{1}$ and R~Hoffmann-Vogel$^{1,2}$}

\address{$^1$ Physikalisches Institut, Karlsruhe Institute of Technology (KIT), D-76131 Karlsruhe, Germany}

\address{$^2$ Institut f\"ur Angewandte Physik, Karlsruhe Institute of Technology (KIT), D-76131 Karlsruhe, Germany}

\ead{r.hoffmann@kit.edu}

\begin{abstract}We study the structure and the electronic properties of Au nanocontacts created by controlled electromigration of thin film devices, a method frequently used to contact molecules. In contrast to electromigration testing, a current is applied in a cyclic fashion and during each cycle the resistance increase of the metal upon heating is used to avoid thermal runaway. In this way, nanometer sized-gaps are obtained. The thin film devices with an optimized structure at the origin of the electromigration process are made by shadow evaporation without contamination by organic materials. Defining rounded edges and a thinner area in the center of the device allow to pre-determine the location where the electromigration takes place. Scanning force microscopy images of the pristine Au film and electromigrated contact show its grainy structure. Through electromigration, a $1.5 \,\mu$m-wide slit is formed, with extensions only on the anode side that had previously not been observed in narrower structures. It is discussed whether this could be explained by asymmetric heating of both electrodes. New grains are formed in the slit and on the extensions on both, the anode and the cathode side. The smaller structures inside the slit lead to an electrode distance below 150 nm. Kelvin probe force microscopy images show a local work function difference with fluctuations of $70$ mV on the metal before electromigration. Between the electrodes, disconnected through electromigration, a work function difference of $3.2$ V is observed due to local charging from contact to the tip. Some of the grains newly formed by electromigration are electrically disconnected from the electrodes.
\end{abstract}

\maketitle

\section{Introduction}

An intriguing idea is to replace semiconductor functional elements in computers by molecules \cite{feynman59t1,aviram74p1}, because the function of the device could then be determined by the chemical design with a multitude of properties accessible \cite{joachim00p1,li15p1}. With this approach, it is essential to know how the electrodes connect to the molecule, i.e. the structure and the electronic properties of the metal/molecule interface and its environment, see e.g. \cite{moth09p1}. Concerning the structure, scanning probe microscopy is well-known as an ideal method to offer structural information even if the substrate is insulating \cite{meyerb1}. Concerning the electronic properties, in particular the chemical potential governs electronic transport. Scanning probe microscopy, using conductive scanning force microscopy \cite{mativetsky14p1} or the Kelvin method \cite{nonnenmacher91p1},  allows to locally determine the work function and the chemical potential even with atomic resolution \cite{sadewasser09p1,perez16p1}. For macroscopic studies, it has been shown how to reconstruct the local surface potential from the measured Kelvin voltage by using assumptions of the point-spread function of tip and cantilever \cite{cohen13p1}. For molecular electronics studies, it is important to know the local electric field in the gap between the electrodes in order to understand whether gating will be efficient for this particular device structure \cite{datta09p1}.

To apply Kelvin methods to molecule-metal functional devices, a device fabrication method allowing scanning probe microscopy studies is needed. The nanocontacts should be as flat as possible to simplify imaging. To avoid a possible influence of contamination by molecules, a fabrication process avoiding the contact with organic chemicals and air should be chosen \cite{stoeffler11p1}. One of the main techniques proposed to this end is the controlled electromigration technique. A device is thinned by subjecting it to a current large enough to induce electromigration \cite{park1999,park2000,durkan2000,schirm13p1}. Monitoring the resistance of the contact and taking it as a measure of the contact's temperature helps to avoid thermal runaway and melting of the contact \cite{shih2003,esen2005,strachan05p1,hoffmann2008}. Using controlled electromigration, parallel fabrication of several contacts on one surface is possible \cite{johnston2007}. A gate-electrode could be fixed on the substrate. If scanning probe microscopy is used, the scanning tip could be viewed as a gate electrode. This yields the advantage that the gate is local and movable. Scanning probe microscopy images of the devices have shown how the grainy structure of the metal alters under the influence of the current \cite{stoeffler12p1,girod12p1,yagi15p1}.

Here, we study the electromigration process as part of the device fabrication for molecular electronics and show that Kelvin studies allow to determine the local connectivity and electrostatic field distribution of the device. We fabricate optimized Au thin film devices for electromigration with a shadow mask without contamination by organic chemicals to obtain a reference of the bare metal electrodes. We control the results of our fabrication process including electromigration using scanning force microscopy. Due to an optimized structure of the thin film device, the electromigration takes place at the center of the device. This is achieved by avoiding sharp edges and by creating a thinner half-shadow area in the center of the device. With such samples, it is possible to apply the controlled electromigration method successfully even though the sample is $200 \,\mu$m wide. Electromigration leads to the formation of a slit of $1,5 \,\mu$m width with the distance between electrodes being at least 10 times less due to smaller metallic structures within this slit. The slit forms extensions parallel to the direction of the electron flow only on the anode possibly due to asymmetric heating. We observe additional grains inside the slit as well as on its edges on the cathode as well as the anode side. Grains are also visible in Kelvin probe force microscopy on the thin film before electromigration, but only lead to small variations of the local work function difference of up to $70$ mV. Charging the disconnected electrodes after electromigration leads to a large electrochemical potential difference of $3.2$ V detected in Kelvin probe force microscopy images. Part of the additional grains in the slit are electrically disconnected from the electrodes.

\section{Experimental Methods}

The thermal conductivity of the substrate is an important parameter for successful controlled electromigration \cite{kiessig14p1,arnold15u1}. We use sapphire as a substrate due to its large heat conductivity ($30.3$ W/(m$\cdot$K)) perpendicular to the c-axis and $32.5$ W/(m$\cdot$K) parallel to the c-axis \cite{dobrovinskaya09b}. The sample was covered with Microposit varnish, cut by a diamond saw, rinsed using isopropanol, acetone, ethanol and distilled water and subsequently dried by heating in air. Then the substrate was subject to an oxygen plasma to remove organic remnants. Remaining impurities visible in an optical microscope were removed using a cotton bud. The Au thin film was nanostructured by shadow evaporation. The mask had a shape of an elongated rectangle of width
$200 \,\mu$m and length $4$ mm with four macroscopic contacts and rounded edges (Fig. 1 a)) to avoid current divergences at edges causing electromigration on the macroscopic contacts rather than - as intended - in the center of the device. In the center of the mask, we mounted a $80 \,\mu$m thin Cu wire to induce electromigration at this position. Au of $28$ nm thickness and for a test sample $200$ nm thickness was deposited by sputter-deposition.

\begin{figure}[t]
    \begin{center}
    \includegraphics[width=8cm]{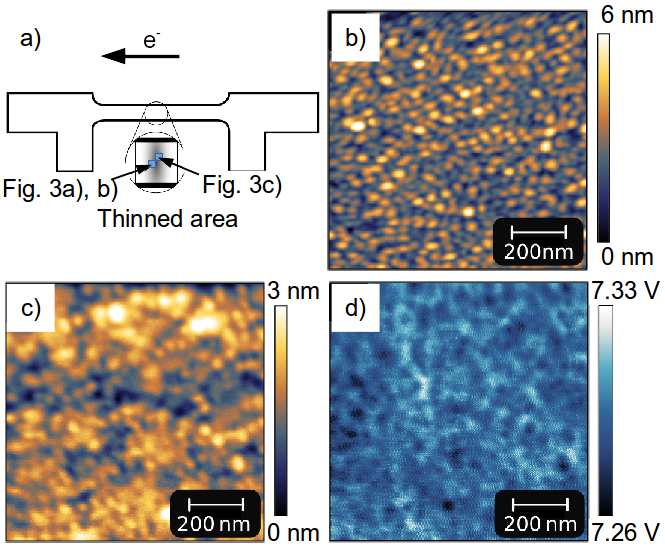}
    \caption
    {a) Mask structure used for shadow evaporation for determining the structure of the electromigrated devices. The area shows the direction of the electron flow, as also in the subsequent figures. The positions, where the images shown in Fig. 3 were obtained are marked by arrows. b) SFM image obtained before electromigration of the area thinned by the half-shadow of the wire spanned over the mask during sputtering. Frequency shift $\Delta f= -35$ Hz. c) Topography and d) local work function difference measured on the 28 nm-thick Au film.  Frequency shift $\Delta f= -28$ Hz.}
    \end{center}
    \label{fig_1}
\end{figure}

For imaging the structure of the device, we used three different scanning force microscopes (SFM). For large-scale measurements in air, we employed a home-built scanning force microscope with a scan range of $400\,\mu$m$\times 400\,\mu$m described in more detail in \cite{ziegler}. This SFM, called large-scale SFM in the following, is operated in the static mode using a cantilever with a force constant of $6$ N/m. In addition we characterize the surface structures using a variable temperature ultra high-vacuum SFM from Omicron, Taunusstein, Germany (VT-AFM) equipped with a Nanonis phase-locked loop electronics (Specs, Z\"urich). The measurements in this instrument, called vacuum SFM in the following, were carried out in the dynamic frequency modulation mode using a Pt-Ir coated Si tip attached to a cantilever of a resonance frequency of $304$ kHz, a force constant of $37$ N/m and an oscillation amplitude of $7$ nm. In this instrument we also performed Kelvin probe force microscopy measurements by applying an oscillating voltage to the tip (frequency of 266 Hz). We nullified the detected signal of this frequency using a lock-in amplifier by applying a voltage to the tip in order to find the work function difference between tip and sample. The third SFM was a Nanosurf easyscan 2 operated in air in the static mode with cantilevers of force constants between $0.02$ N/m and $0.77$ N/m.

\subsection{Structure and local work function before electromigration}

Before electromigration, SFM images using the large-scale SFM showed that the intended thickness of $200$ nm is indeed obtained in the area undisturbed by the wire. In the center of the half-shadow caused by the wire, the film is only $56$ nm thick. We conclude that we can expect a factor of 1/4 of the film thickness in the half-shadow area for this method. We estimate that the current density in the half-shadow area was increased by a factor of four, causing electromigration to start in this area. The lateral extension of the half-shadow area was approximately $100 \,\mu$m, only slightly larger than the diameter of the wire, $80 \,\mu$m. A measurement using a profilometer (Ambios technology, Santa Cruz, California) confirms this measurement within the limit of experimental errors by showing a width of $180 \,\mu$m of the wire and a thickness of $42$ nm. On the undisturbed area of the wire, we obtain a height of $260$ nm of the Au film using the profilometer.

We investigated the microstructure of the Au film using the vacuum SFM (Fig. 1 b)). In the half-shadow area, we observe Au grains of a diameter between $40$ and $80$ nm and a height between $1$ and $4$ nm. These results are in good agreement with results obtained on the area far from the half-shadow, with results on mica \cite{arnold15u1} and on SiO$_x$ \cite{stoeffler12p1}. The local work-function differences obtained using Kelvin probe force microscopy show only small local variations of the work function of up to $70$ mV.

\subsection{Electromigration electrical characterization}

After structural characterization, we subjected the Au thin film to cyclic current ramps in two-terminal configuration. This is different from previously used methods, where mainly voltage ramps were used \cite{esen2005,hoffmann2008}. It had been proposed that avoiding disturbances due to the leads is advantageous for the electromigration process and that using a four-probe set-up helps to know the resistance of the forming contact and to avoid an influence of the leads \cite{wu07p1}. Here, we intend to obtain a similar independence of the lead resistance by using current-control rather than voltage-control. Within each cycle, after each current step, we measure the resistance and compare the measured value with a pre-chosen resistance threshold. We continue to increase the current until the threshold is reached. We then automatically reduce the current to end the cycle and increase the resistance threshold for the next cycle. After that we start the next cycle by increasing the current again.

Electromigration starts with a regime where the resistance reversibly increases due to Joule heating (Fig. 2a)). A convex regime follows where the resistance starts to change permanently followed by an undercut. The overall shape of the undercut can in general be explained by a model where the lead resistance $R_L$ (i.e. the resistance of most of the thin film device excluding the forming contact) as well as the electric power dissipated in the contact $P^*$ are taken to be constant as a function of current $I$ such that the total voltage becomes \cite{strachan05p1}

\begin{equation}
 U= R_L I + P^*/I
\end{equation}

Values to the right of the curve correspond to excessive dissipated power and hence overheating of the contact. Here, we show such a model curve (green line in Fig. 2 a)), however, the electrical characteristics of the device do not follow this curve. It has been shown previously that the model can be refined by assuming that upon thinning, parts of the device that were counted as part of the forming contact at the start of the electromigration process, must be accounted for as leads as the contact forms and the current becomes more and more locally confined \cite{stoeffler14p1}. This can be done taking a second model with larger $R_L$ and smaller $P^*$ as the heated volume shrinks. Possibly, the curve could be represented by taking 5 or 6 such models. Independently of this peculiarity, the power dissipated in the contact is limited, and neither the electrical data nor the structural analysis discussed below shows a sign of global overheating. This is particularly surprising due to the large width of the samples ($200 \,\mu$m) compared to previously published results where the samples were only several 500 nm wide at the most \cite{hoffmann2008,stoeffler12p1}.

\begin{figure}[t]
    \begin{center}
    \includegraphics[width=8cm]{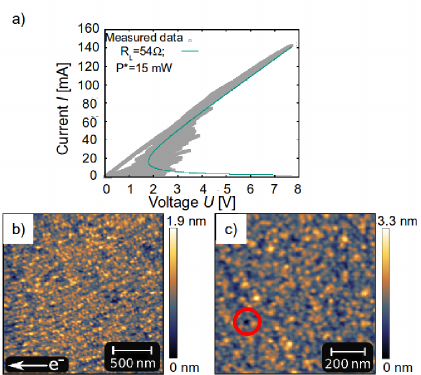}
    \caption
    {a) Electromigration data: measured voltage and set current during computer-controlled electromigration. The green line shows an example curve following equation 1. SFM image of the area at b) the anode and c) the cathode of the device thinned by electromigration about $100\pm 70 \,\mu$m from the half-shadow area. In c) the area of a void is marked with a red circle. b) Frequency shift $\Delta f= -24$ Hz. c) Frequency shift $\Delta f= -14$ Hz.}
    \end{center}
    \label{fig_2}
\end{figure}

\subsection{Sample structure and local work function after electromigration}

Optical images taken after electromigration show no topographical changes of the sample. In scanning force microscopy images on positions far from the electromigrated area on the device and on the anode, nearly no differences to the measurements before electromigration are observed (Fig. 2 b) and c)). The corrugation is approximately $1$ nm. On the cathode (Fig. 2 c)), few voids are visible, as marked by a red circle.

For SFM measurements in air in the static mode images show the anode and the cathode side of the slit that was formed (Fig. 3) in agreement with our previous results~\cite{stoeffler12p1}. It is surprising that in spite of the large width ($200 \,\mu$m) of the samples used here compared to previous ones ($500$ nm), a relatively uniform slit of a width of $1.5 \,\mu$m is formed. At a number of positions, the slit forms extensions towards the anode side that have previously not been observed. The edges of the slit are decorated with grains of the size of up to $200$ nm at the cathode and the anode side. Due to their topographically elevated position with respect to the sample plane and their location on the edge of the slit, we assume that they are newly formed grains. The circular shape of the newly formed grains points to local overheating and melting of the Au layer as had been proposed previously \cite{stoeffler12p1}. Due to the small size of the newly formed grains of only 100 to 200 nm, it is clear that the heating remained locally restricted to an area of similar length scale. The newly formed grains are observed also on the edges of the slit-extensions indicating that they were formed after the formation of the extensions. This also means that the two electrodes must have been electrically connected after the formation of the extensions, by metallic contact or by tunneling. Within the slit, smaller grains of up to about $100$ nm in size are observed showing that the width of the slit is not identical with the distance of the electrodes at the gap. The gap distance is estimated for our images to be smaller than 150 nm.

\begin{figure}[t]
    \begin{center}
    \includegraphics[width=8cm]{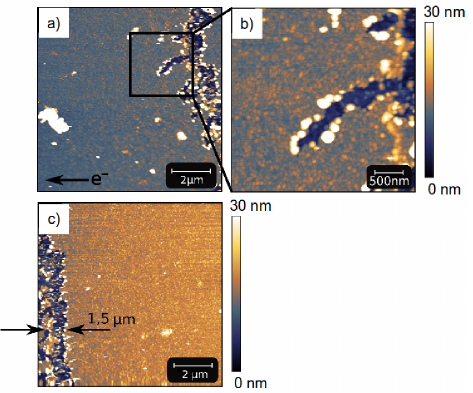}
    \caption
    {SFM image of the area affected by electromigration a) Overview of the anode side. b) Area shown by a black square in a) enlarged. c) Slit formed and overview of the cathode side.}
    \end{center}
    \label{fig_3}
\end{figure}

We additionally investigated the area of the formed contact using the vacuum-SFM (Fig. 4 a)). The metallic layer on the sample was not electrically contacted, only the substrate back plate. For this sample, the two electrodes were charged accidentally. This turned out to be useful for determining the position of the slit. Charging the electrodes could be done on purpose in similar experiments allowing for faster and more reliable positioning of the measuring probe. The large height difference of $55$ nm between anode and cathode observed in the SFM image at only $-3$ Hz frequency shift is due to additional electrostatic forces occurring on the anode side. On the upper and lower part of the slit we find a similar width compared to Fig. 3 of about 1 $\mu$m, whereas a reduced distance of the electrodes forming the gap is observed in the center of the image. The gap region had an apparent width of $270$ nm in this measurement and appears enlarged due to the additional electrostatic forces.

\begin{figure}[t]
    \begin{center}
    \includegraphics[width=8cm]{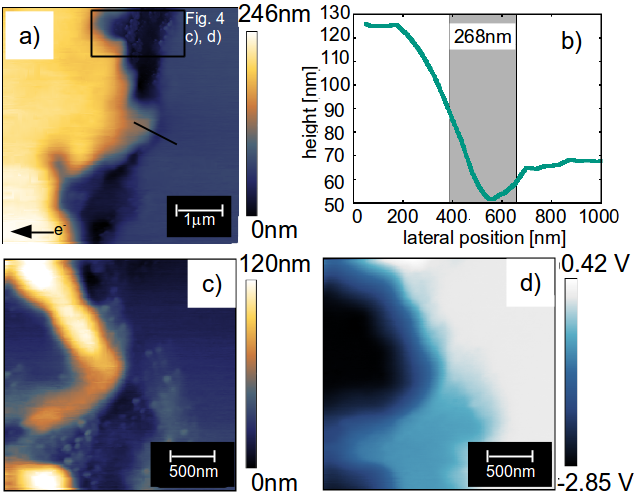}
    \caption
    {SFM image of the area affected by electromigration a) Overview image. A black line shows the position of the line profile shown in b) and a black frame shows the position of a close-up image shown in c). For a) the frequency shift was $\Delta f= -3$ Hz. b) Line profile along the black line in part a). SFM image of the area affected by electromigration. For c) the frequency shift was $\Delta f= -4$ Hz. d) Kelvin probe force microscopy image measured in parallel to c).}
    \end{center}
    \label{fig_4}
\end{figure}

In order to investigate this area in more detail, we used Kelvin probe force microscopy (Fig. 4 c) and d)). By the Kelvin method, we measure the potential difference $\Delta V_{\mbox{\scriptsize Kelvin}}$ caused by a difference in the electrochemical potential of the surface $\eta_{\mbox{\scriptsize surface}}$ with respect to the tip $\eta_{\mbox{\scriptsize tip}}$

\begin{equation}
 e\Delta V_{\mbox{\scriptsize Kelvin}}=\eta_{\mbox{\scriptsize surface}}-\eta_{\mbox{\scriptsize tip}}
\end{equation}

Measuring a different $\Delta V_{\mbox{\scriptsize Kelvin}}$ on two different parts of the sample with the same tip can thus be translated into a electrochemical potential difference between the two parts of the sample, where the role of the reference, the tip, becomes obsolete. The electrochemical potential comprises the chemical potential $\mu$ and the electrostatic potential $eV_0$, i.e. $\eta=\mu+eV_0$. Since both electrodes are made from the same material, we assume the chemical potential difference $\Delta \mu$ between them to be near zero, and the values in Fig. 4 to be determined by voltage differences $eV$.

Indeed, as anticipated in the discussion of Fig. 4 a), a strong variation of the Kelvin potential difference $\Delta V_{\mbox{\scriptsize Kelvin}}$ of the anode compared to the cathode of $3.2$ V is observed (Fig. 4 d)). This large Kelvin potential difference shows that the two contacts are definitely electrically disconnected after electromigration and that tunneling between the two electrodes is unlikely. These results could be used to reconstruct the surface potential, and if a molecule were placed between the contacts and one could determine whether gating could be efficient in this configuration \cite{datta09p1}. The large potential difference leads to strong electrostatic forces originating from the charged surface and makes it difficult to obtain high resolution. The distance between the electrodes, $270$ nm, and the electrostatic potential difference yields a simple estimate of the electrostatic field in the gap, $10$ mV/nm. This value is small compared with other electrostatic fields occurring at the atomic scale, e.g. the electrostatic field differences at the atomic scale on the NaCl (001) surface \cite{steurer15p1}.

\section{Discussion}

\subsection{Overall comparison with previous results}

The overall microstructure is in good agreement with previous results~\cite{stoeffler12p1}. Previously, we had investigated the voltage dependence of the resistance of Au contacts formed by electromigration in vacuum and found a negative slope of the resistance as a function of voltage \cite{stoeffler14p1}. This negative slope was attributed to a nanocontact of a large area in the tunneling regime and a change of the distance of the two electrodes of the nanocontact due to either thermal expansion, when the contact is heated resistively, or electrostatic forces due to the applied voltage. Although tunneling is unlikely here due to the large distance of the electrodes, the structure of the sample is in good agreement with the structure proposed in \cite{stoeffler14p1}: There is a large area of the slit, because its width is $200 \,\mu$m and within the slit, it is easily possible to have many tunneling contacts in parallel at an earlier stage of the electromigration process. Also the assumption of a grainy structure of the sample taken in Ref. \cite{stoeffler14p1} is well supported by the measurements shown here.

\subsection{Formation of extensions}

Voids in the longitudinal direction had been observed before, e.g. in an aluminum line \cite{shingubara91p1}. The appearance of such longitudinal voids was associated with enhanced electromigration stability, because a slit extending in the direction perpendicular to the electron flow was not formed in that case. Here, in contrast, we observe both, a slit crossing the path of the current as well as voids forming in the longitudinal direction.

One possible explanation is that these extensions could have been caused by a broadened temperature distribution. It has been discussed previously that the width of the temperature distribution can change in the course of the electromigration thinning process \cite{stoeffler12p1}. As the contact thins, also the heated region becomes smaller because the electron flow is distributed over a smaller width of the thin film and less power is needed to heat this smaller volume. This process usually becomes evident from the electromigration I(U) data (Fig. 2) and is active also here as discussed above. In this view the extensions might be regarded as the result of a broad temperature distribution occurring at the start of the electromigration process. However, if this were the only process, we would expect that also the area between the extensions were affected by the broad temperature distribution. For example such a process has been observed in electromigrated thin film alloys \cite{kozlova13p1} where evenly distributed voids were observed. In alloys, the formation of voids and islands can be explained by inhomogeneities in the composition of the thin film. In fact, the presence of the extensions in a pure Au film points to a temperature distribution that has inhomogeneities in the direction parallel to the slit (i.e. perpendicular to the electron flow).

One consideration for the origin of these inhomogeneities could be inhomogeneities in the current flow - by interference and/or local defects. However, the size of the slit is much longer than the phase coherence length of electrons in Au \cite{hasegawa93p1,dumpich91p1}. The presence of the extensions points to a combined process, where the current is guided within a certain area due to enhanced electromigration in the same area. Curiously, the process happens to occur only at the anode side of the slit, pointing to enhanced heating on only this side of the slight. Recently, a new mechanism of electromigration failure by waterwheel phonon modes has been described, where electric failure occurs at low voltages deterministically \cite{dundas09p1,lue10p1}. This mechanism is related to selective heating of only one electrode \cite{lue15p1,tsutsui12p1,lee13p1}.

\begin{figure}[t]
    \begin{center}
    \includegraphics[width=8cm]{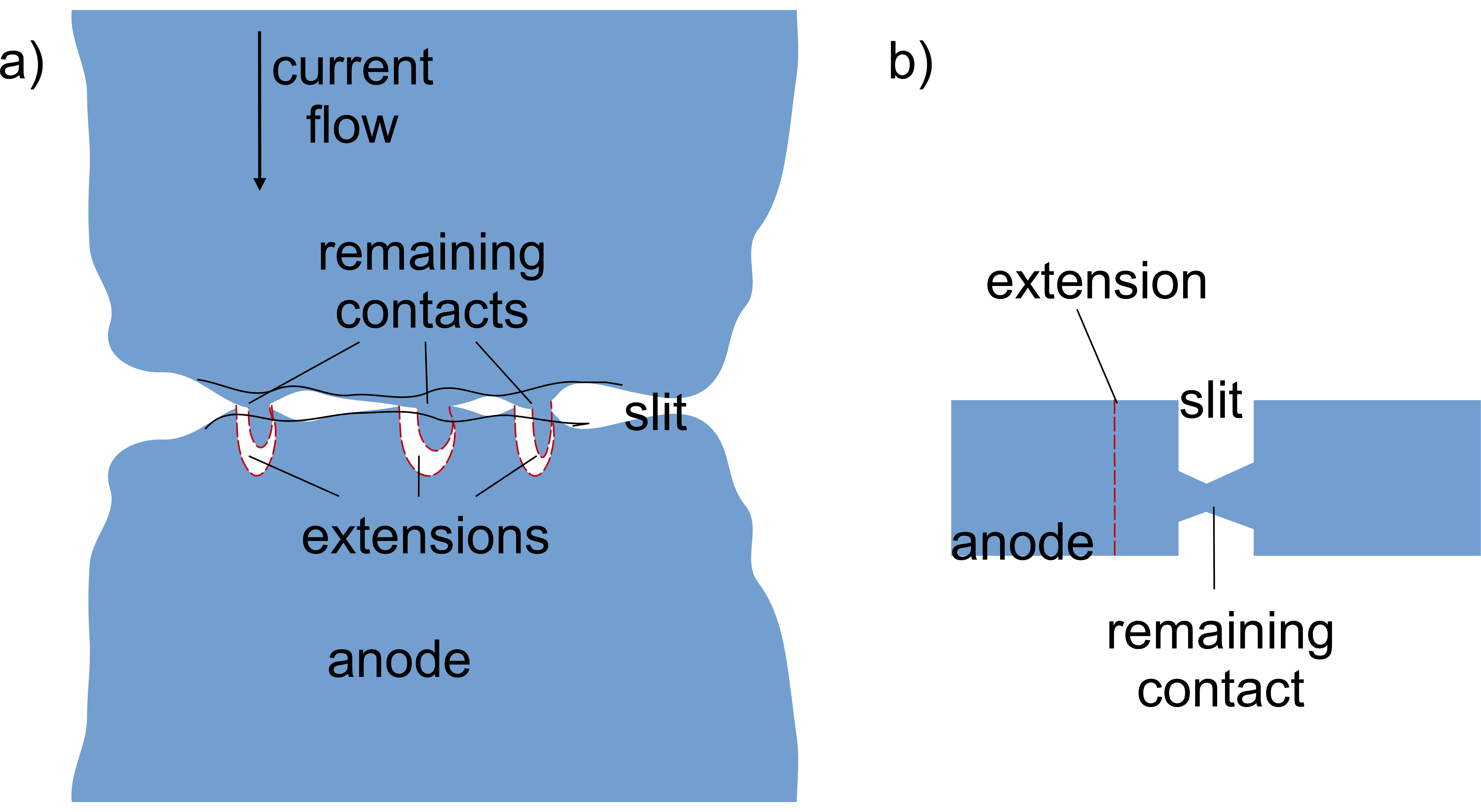}
    \caption
    {a) Topview and b) sideview of the possible development of the structure: At the start of the electromigration process, the slit is formed across the full width of the thin film with only a few contacts remaining. These are first thinned by electromigration to become nanocontacts and extensions form along the slit. At the final breakage, additional power is dissipated in these nanocontacts.}
    \end{center}
    \label{fig_5}
\end{figure}

The local guidance of the electron flow could arise naturally as the contact breaks up into several nanocontacts during the formation of the slit. For geometrical reasons most of the material transport causing slit formation occurs at the start of the electromigration process, when the resistance increases by a factor 2 or 3 \cite{stoeffler12p1}. At this point, the contact is still electrically connected and it is reasonable to assume, for a structure of the width of $200 \,\mu$m, that the electrical connections occur at several positions. As these are thinned, they eventually reach the ballistic regime in which energetic electrons are injected into the anode, and thermalized primarily on that side. In the ballistic regime, in addition to standard electron-hole asymmetries responsible for electromigration effects, additional such asymmetries occur due to waterwheel, i.e. rotational, phonon modes.

In this picture, the position of the extensions would reflect the position of the remaining contacts formed at the start of the electromigration process, see Fig. 5. We expect that these form at positions where either the local activation energy for diffusion is higher, e.g. due to a more defect-free area, or, where the local cooling is better due to a better thermal contact to the environment. The number of extensions per unit length gives an indication on how frequently such positions occur in the film.

To confirm that selective heating of one electrode really is responsible for the formation of the extensions, it should be carefully excluded that such extensions could be caused by classical current focusing as the slit forms. Hot spot formation by current focusing was indeed found to lead to slit formation, but the slits started forming in the direction perpendicular to the electron flow and later extended in the direction {\it opposite to, i.e. upstream of} the electron flow \cite{mahadevan99p1} whereas here we observe the formation of extensions in the direction {\it with, i.e. downstream with} the electron flow. This classical upstream direction of slit formation is directly related to the downstream material transport that is expected for metal electromigration. The overall downstream material flow is in agreement with additional voids observed here far from the slit on the cathode side, see Fig.~2c. Here, we regard the formation of extensions as related to new type of electron-hole asymmetries and selective heating of electrodes as they occur from waterwheel modes \cite{lue15p1}.

\subsection{Additional grains decorating the slit}

We have already concluded above that the small additional grains must have been formed after the formation of the slit and its extensions and that their shaped points to local melting. Since we measure at constant current, in the moment where the contact breaks, the total power sum dissipated in the remaining nanocontacts $P=RI^2$ increases until breakage occurs. This is in contrast to previous observations of local melting where electromigration was performed at constant voltage \cite{stoeffler12p1}.

Grains in the gap were predicted first from electric measurements of gated nanocontacts, where Coulomb blockade behavior was observed, see, e.g. \cite{houck05p1}. Those grains should only have a size of about 20 atoms - much smaller than the largest ones observed here. Additional smaller ones could be below the resolution limit of Fig. 3. The question arises, whether the grains observed here are electrically connected to the leads. In Fig. 4 c) and d) one can see a region within the slit that comprises such grains, as clearly seen in the topographical image, Fig. 4 c). In the Kelvin image of the local electrostatic potential, Fig. 4 d) part of the grains are not visible even if the image is analyzed closely. We believe that this occurs because the potential of the islands varies smoothly as the electrostatic field in that region. The electrostatic field in that region originates from the cathode and the anode of the slit. If the grains were electrically connected to one of the two electrodes, we would expect that that grain should have a similar electrostatic potential as that electrode. For some grains this is clearly not the case; these grains must be electrically disconnected. Smaller grains could cause Coulomb blockade effects as have been observed.

\section{Conclusions}

Combined topographic and Kelvin scanning force microscopy studies allow us to understand the relation between topography and electronic properties in electromigrated nanocontacts. We show that electromigration thinning down to nanocontacts is possible even for thin film bridges as wide as $200 \,\mu$m. A thinner area in the center and rounded edges allow to pre-determine the location of the nanocontact where a $1.5 \,\mu$m-wide slit is formed. For such a wide thin film as opposed to a narrower structure, extensions of the slit in the direction perpendicular to the electron flow are observed. These point to local inhomogeneities in the electron flow and to stronger heating of the anode side compared to the cathode side since the direction of material transport by electromigration is expected to be with the electron flow. Additional grains are formed when the contact finally breaks due to local melting. Some of these grains are electrically disconnected from both the anode and the cathode. 

\ack
This work was supported by the ERC Starting Grant NANOCONTACTS (No. 239838) and by the ministry for science, research and arts, Baden-W\"urttemberg, through the Brigitte-Schlieben-Lange program. We thank Th.~Schimmel and A.~Hasenfu{\ss} for help with measurements using the large scale scanning force microscope home-built in the Schimmel research group and Ch.~S\"urgers for help with sample preparation.

\end{document}